\documentclass[]{aastex}
\usepackage{emulateapj5}
\usepackage{epsfig}
\newcommand{\LCDM}{$\Lambda$CDM}
\newcommand{\WMAP}{{\it WMAP}}

\newcommand{\hmpc}{{{\rm h}^{-1}\, {\rm Mpc}}}

\newenvironment{inlinefigure}{
\def\@captype{figure}
\noindent\begin{minipage}{0.999\linewidth}\begin{center}}
{\end{center}\end{minipage}\smallskip}

\begin{document}
\slugcomment{{\em submitted to Astrophysical Journal Letters}}

\shorttitle{Cosmic Variance in GOODS}

\shortauthors{Somerville et al.} 

\title{Cosmic Variance in the Great Observatories Origins Deep Survey}
%\title{A Cosmic Variance Cheatsheet for the GOODS Survey}

%-------------------------------------------------------------
\author{Rachel S. Somerville\altaffilmark{1}, Kyoungsoo Lee\altaffilmark{2},
Henry C. Ferguson\altaffilmark{1}, 
Jonathan P. Gardner\altaffilmark{3}, Leonidas
A. Moustakas\altaffilmark{1}, Mauro Giavalisco\altaffilmark{1}}

\altaffiltext{1}{Space Telescope Science Institute, Baltimore MD 21218; 
somerville,ferguson,leonidas,mauro@stsci.edu}

\altaffiltext{2}{Department of Physics and Astronomy, Johns Hopkins
University, Baltimore, MD 21218; soolee@stsci.edu}

\altaffiltext{3}{Laboratory for Astronomy and Solar Physics, Code 681, Goddard Space Flight Center, Greenbelt MD 20771; jonathan.p.gardner@nasa.gov
}

\begin{abstract}
Cosmic variance is the uncertainty in observational estimates of the
volume density of extragalactic objects such as galaxies or quasars
arising from the underlying large-scale density fluctuations. This is
often a significant source of uncertainty, especially in deep galaxy
surveys, which tend to cover relatively small areas. We present
estimates of the relative cosmic variance for one-point statistics
(i.e. number densities) for typical scales and volumes sampled by the
Great Observatories Origins Deep Survey (GOODS). We use two
approaches: for objects with a known two-point correlation function
that is well-approximated by a power law, one can use the standard
analytic formalism to calculate the cosmic variance (in excess of shot
noise). We use this approach to estimate the cosmic variance for
several populations that are being studied in the GOODS program:
Extremely Red Objects (ERO) at $z\sim1$, and Lyman Break Galaxies
(LBG) at $z\sim3$ and $z\sim4$, using clustering information for
similar populations in the literature. For populations with unknown
clustering, one can use predictions from Cold Dark Matter theory to
obtain a rough estimate of the variance as a function of number
density. We present a convenient plot which allows one to use this
approach to read off the cosmic variance for a population with a known
mean redshift and estimated number density. We conclude that for the
volumes sampled by GOODS, cosmic variance is a significant source of
uncertainty for strongly clustered objects ($\sim 40$-60\% for EROs)
and less serious for less clustered objects, $\sim 10$-20\% for LBGs.
 
\end{abstract}

\subjectheadings{galaxies: statistics --- large scale structure of universe}

%=======================
% 1
\section{Introduction}
\label{sec:intro}
%=======================

The number density of observed extragalactic populations in the
Universe is a fundamental property which may hold clues to the nature
of the objects. However, observational estimates of the number density
of any clustered population are plagued by uncertainty due to
\emph{cosmic variance}, the field-to-field variation (in excess of
Poisson shot noise) due to large scale structure. Clearly, if one can
sample a volume that is very large compared with the intrinsic
clustering scale of the objects in question, cosmic variance will be
insignificant. In practice, especially in high-redshift studies, the
volumes sampled are small enough that cosmic variance is often a
significant source of uncertainty. Perhaps the majority of published
cosmological number densities and related quantities (e.g. luminosity
functions, integrated luminosity densities, etc.) do not properly
account for cosmic variance in their quoted error budgets.

Cosmic variance has frequently been invoked as a motivation for
carrying out deep pencil-beam surveys along multiple sightlines.
While the term ``cosmic variance'' is generally understood, the effects
are of course dependent on the clustering properties of the sources of
interest, a fact that is often lost in discussions of deep survey
strategy. With the availability of the GOODS data, it seems
appropriate to cast this variance in practical terms, calculating
explicitly the expected uncertainties due to clustering for various
source populations under study.  A simple exposition of this cosmic
variance may be useful both to researchers using the GOODS data and to
those planning future studies.

The mean $\langle N\rangle$ and variance $\langle N^2\rangle$ are the
first and second moments of the probability distribution function
$P_N(V)$, which represents the probability of counting $N$ objects
within a volume $V$. We define the \emph{relative} cosmic variance:
\begin{equation}
\sigma^2_{v} \equiv \frac{\langle N^2\rangle -\langle
N\rangle^2}{\langle N \rangle^2} - \frac{1}{\langle N\rangle} \, .
\end{equation}
Note that the last term is the usual correction for Poisson shot
noise, which for the samples considered here will typically be
negligible.  In any case, it is relatively straightforward to perform
this correction, so we do not discuss this term further.  In the
general hierarchical scenario of structure formation, in which density
perturbations grow via gravitational instability, $P_N$ is expected to
have non-zero higher moments (e.g. skewness and kurtosis).  For a
detailed and general treatment of the cosmic error, see \citet{csjc},
\citet{szapudi:99,scjc}, and references therein.  Here, we concentrate
solely on one-point statistics (i.e. counts in cells) and do not
address the cosmic error with respect to two-point or higher order
statistics such as correlation functions. We defer treatment of these
issues to future works.

For a population with a known two-point correlation function $\xi(r)$,
it is straightforward to calculate the cosmic variance as a function
of cell radius $R$ or equivalently, cell volume $V$ \citep[see
e.g.][or Section~\protect\ref{sec:pl} below]{peebles:80}. There are,
however, several potential practical difficulties with this simple
approach. While the correlation function of galaxies is typically
well-approximated by a power law in the strongly non-linear regime ($r
\la 10$-15 Mpc), on larger scales, in the linear regime, the
correlation function is expected to deviate from the power-law slope
measured on smaller scales. Also, estimating the correlation function
(especially its slope) of an observed population is more difficult
than estimating the number density, so often the latter quantity is
known while the former is not. In this situation, we can use the
theory of clustering and bias in the Cold Dark Matter (CDM) paradigm
to estimate the cosmic variance for a population with a known mean
redshift and average comoving number density.

In this \emph{Letter}, we estimate the uncertainty due to cosmic
variance for several populations that have been identified in the
GOODS survey, and present general results based on CDM theory that can
be used to estimate the cosmic variance for populations at $z<6$.
Throughout, we assume cosmological parameters consistent with the
recent analysis of \WMAP\ data \citep{spergel:03}: matter density
$\Omega_m = 0.3$, baryon density $\Omega_b =0.044$, cosmological
constant $\Omega_{\Lambda}=0.70$, Hubble parameter $H_0=70$ km/s/Mpc,
fluctuation amplitude $\sigma_8 = 0.9$, and a scale-free primordial
power spectrum $n_s=1$.

\section{The geometry of GOODS}

\begin{figure*} 
\epsscale{2.0}
%\clearpage
%\begin{figure} 
\plotone{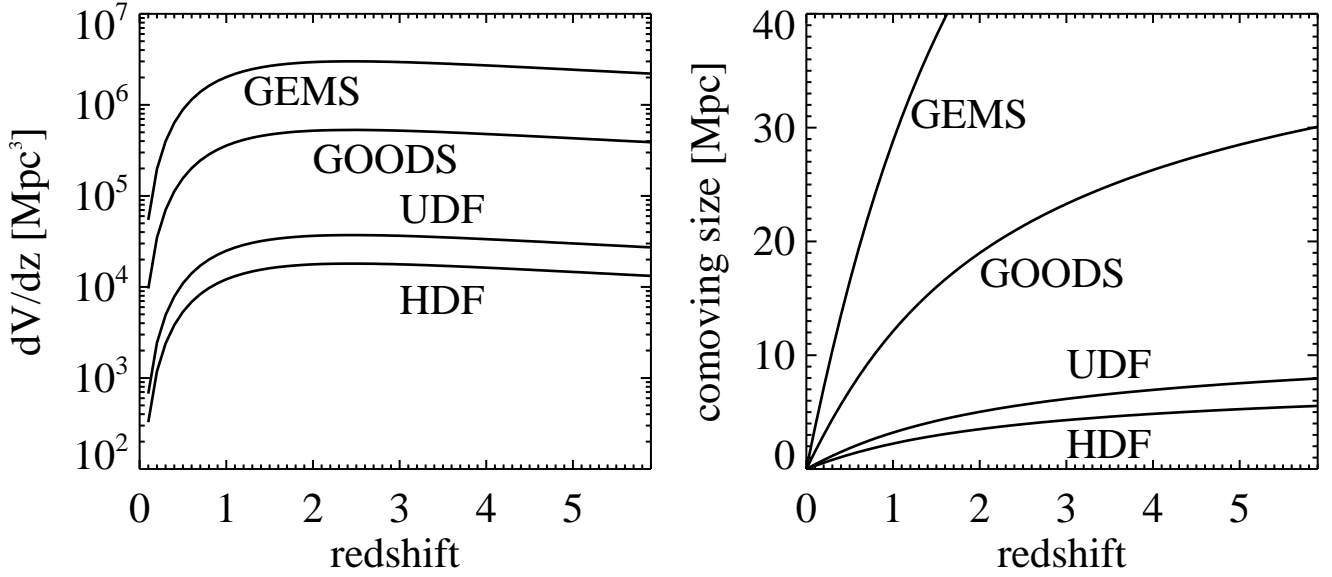}
\caption{\small [left] The comoving volume per unit redshift spanned
by (from bottom to top), the original HDF, the UDF, the GOODS field
and the GEMS field. [right] The transverse comoving size of the HDF,
UDF, GOODS, and GEMS fields.
\label{fig:scale}}
\end{figure*}
%\end{figure}
%\clearpage
The Great Observatories Origins Deep Survey (GOODS) covers two fields,
the Chandra Deep Field South (CDFS) and the Hubble Deep Field North
(HDFN). The CDFS field has dimensions 10'$\times$16', and the GOODS
HDFN field has similar dimensions. For more details and a general
overview of the GOODS program, see \citet{giavalisco:03}. Here, we
treat the case of a single CDFS-sized field.  For widely separated
fields, the cosmic variance goes as $1/N_{\rm field}$, so the variance
will decrease by a factor of two when the second field is included.
In Fig.~\ref{fig:scale}, we show the comoving volume per unit redshift
for several recent, ongoing, and planned deep HST surveys: the
original Hubble Deep Field North \citep{williams}, the GOODS CDFS
field, the GEMS (Galaxy Evolution from Morphology and SEDs; Rix et
al. in prep)
field\footnote{http://www.mpia.de/homes/barden/gems/gems.htm}, and the
planned Ultra Deep Field (UDF)
\footnote{http://www.stsci.edu/science/udf/}. For redshifts $z\ga1$
and $\Delta z \sim 0.5$--1, GOODS samples a volume of a few $\times
10^5\, {\rm Mpc^3}$. Fig.~\ref{fig:scale} also shows the average
transverse size ($L = \sqrt{10'\times 16'} = 12.7'$) of the GOODS
field as a function of redshift, again compared with the original HDF
($L = \sqrt{5.7 \rm arcmin^2} = 2.4'$). 

\section{The Power-Law Model}
\label{sec:pl}

The relative cosmic variance for a population with known two-point
correlation function $\xi(r)$ is given by:
\begin{equation}
\sigma^2_v = \frac{1}{V^2} \int^R_0 \xi(|{\bf r_1}- {\bf r_2}|)\, dV_1 dV_2
\end{equation}
\citep[see, e.g.][p. 234]{peebles:80}. If the correlation function can
be represented by a power-law $\xi(r) = (r_0/r)^\gamma$, then this
expression can be evaluated in closed form:
\begin{equation}
\sigma^2_v = J_2\, (r_0/r)^\gamma
\label{eqn:cv_pl}
\end{equation}
where $J_2 = 72.0/[(3-\gamma)(4-\gamma)(6-\gamma)2^\gamma]$
\citep[][p. 230]{peebles:80}. Assuming spherical cells, the variance
may be equivalently expressed in terms of the cell radius $R$ or the
cell volume $V\equiv 4 \pi R^3/3$. 

\begin{table*}
\begin{center}
\caption{Summary of parameters for representative populations
\label{tab:param}}
\begin{tabular}{lcccccc}
\tableline
object & $\bar{z}$ & mag. limit &
$n$ (h$^3$ Mpc$^{-3}$) & $r_0$ (h$^{-1}$ Mpc) & $\gamma$ & Ref. \\ 
\tableline
ERO & 1.2 & $K_s<19.2$ & $\sim 10^{-3}$ & $12 \pm 3$ & [1.8] & D2001 \\
%\protect\citet{daddi:01} \\ 
ERO & 1.2 & $18 < H < 20.5$ & $(1.0 \pm 0.1) \times 10^{-3}$ & $9.5\pm5$ & [1.8] & M2001\\
%\protect\citet{mccarthy} \\ 
U-drop & 3 & $\mathcal{R}<25.5$ & $4.7 \times 10^{-3}$ & $3.96 \pm 0.29$ & $1.55 \pm 0.15$ & A2003 \\
%\protect\citet{adel:03}\\
B-drop & 4 & $i'<26$ & $1.78 \times 10^{-3}$ & $2.7^{+0.5}_{-0.6}$ & [1.8] & O2001\\
%\protect\citet{ouchi:01}\\
\tableline
\end{tabular}
\tablecomments{The mean redshift, magnitude limit, number density,
correlation length, and correlation function slope for the populations
shown in Fig.~\protect\ref{fig:var_pl}. References are as follows:
D2001 -- \protect\citet{daddi:01}; M2001 -- \protect\citet{mccarthy};
A2003 -- \protect\citet{adel:03}; O2001 -- \protect\citet{ouchi:01}.
Where the correlation function slope is in brackets, this indicates
that the value was assumed in, rather than derived from, the
analysis.}
\end{center}
\end{table*}

For objects with a known correlation function that is well represented
by a power-law, we can simply use Eqn.~\ref{eqn:cv_pl} to compute the
cosmic variance for a given effective volume, as illustrated in
Fig.~\ref{fig:var_pl}. We show $\sigma_v$ as a function of volume, for
three populations with correlation function estimates from the
literature: Extremely Red Objects (EROs) at mean redshift $\bar{z}\sim
1.2$, U-band dropouts at $\bar{z}\sim 3$ (also known as Lyman break
galaxies (LBG)), and B-band dropouts at $\bar{z}\sim 4$.  The
magnitude limit and color selection used for each of these populations
selects objects in a given redshift range, resulting in an effective
volume $V_{\rm eff}$.  Characteristic number densities for each of
these populations, along with correlation function parameters and the
relevant references, are summarized in Table~\ref{tab:param}. For
example, for EROs in the GOODS field, $\sigma_v \sim 0.4-0.6$, while
for the less clustered LBGs, $\sigma_v \sim 0.15-0.2$. Note that we
have assumed here a spherical geometry for the cells, while in fact
for the GOODS survey the cells are very elongated, with the redshift
dimension being much longer (about a factor of ten) than the
transverse dimension in comoving distance units. We have also ignored
the evolution in clustering that occurs over the time interval between
the `back' and the `front' of the cell. It should be noted that, for
two fields with the same volume, the cosmic variance is \emph{smaller}
for an elongated (parallelepiped or cylindrical) field than for a
compact (cubical or spherical) field \citep[see
e.g.][]{newman:02}. This is because an elongated field samples more
independent (uncorrelated) regions. Therefore, the estimates given
here provide an upper bound on the cosmic variance.

%\clearpage
\begin{inlinefigure}
\begin{center}
\resizebox{\textwidth}{!}{\includegraphics{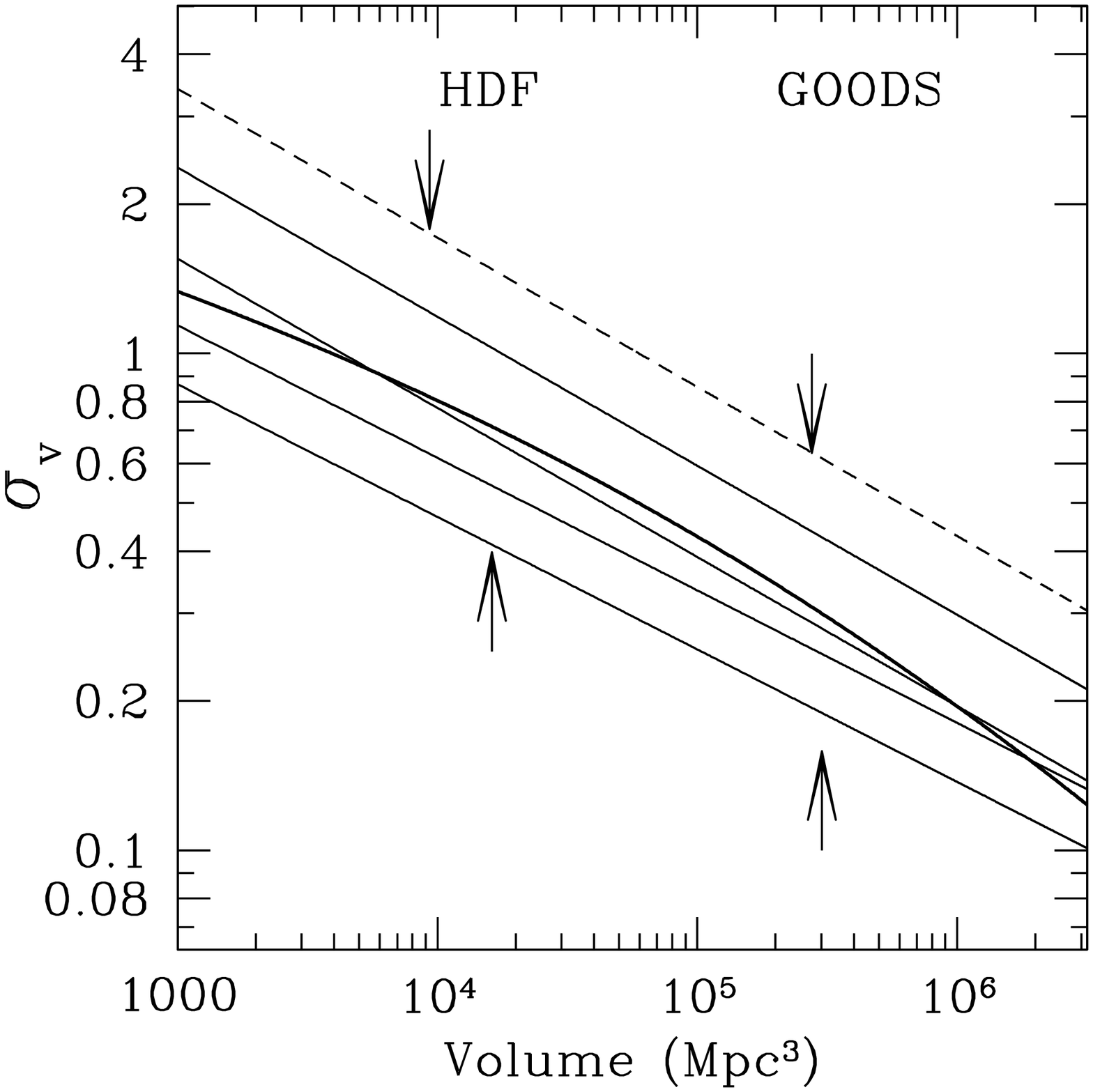}}
\end{center}
\figcaption{\small The square root of the cosmic variance, $\sigma_v$,
is plotted as a function of cell volume $V$. The middle set of lines
are for objects which are as clustered as the dark matter at
$z=0$. The slightly curved line shows $\sigma_{\rm DM}$ from linear
theory, while the straight line is a power law model with $r_0=5
\hmpc$ and $\gamma=1.8$. The topmost solid and dashed lines are for
objects as clustered as EROs at $z=1.2$ (with $r_0=9.0 \hmpc$ and
$r_0=12 \hmpc$, respectively). The second to the bottom line is for
objects that cluster like U-dropouts (LBGs) at $z=3$, and the
bottom-most line is for objects as clustered as B-dropouts at
$z\sim4$. The arrows show representative effective volumes for the
original HDF and GOODS, for EROs (top set) and LBGs (bottom set).
\label{fig:var_pl}}
\end{inlinefigure}

%\clearpage

\section{CDM models}
%\clearpage
\begin{figure*} 
%\begin{figure} 
%\epsscale{2.0}
\plottwo{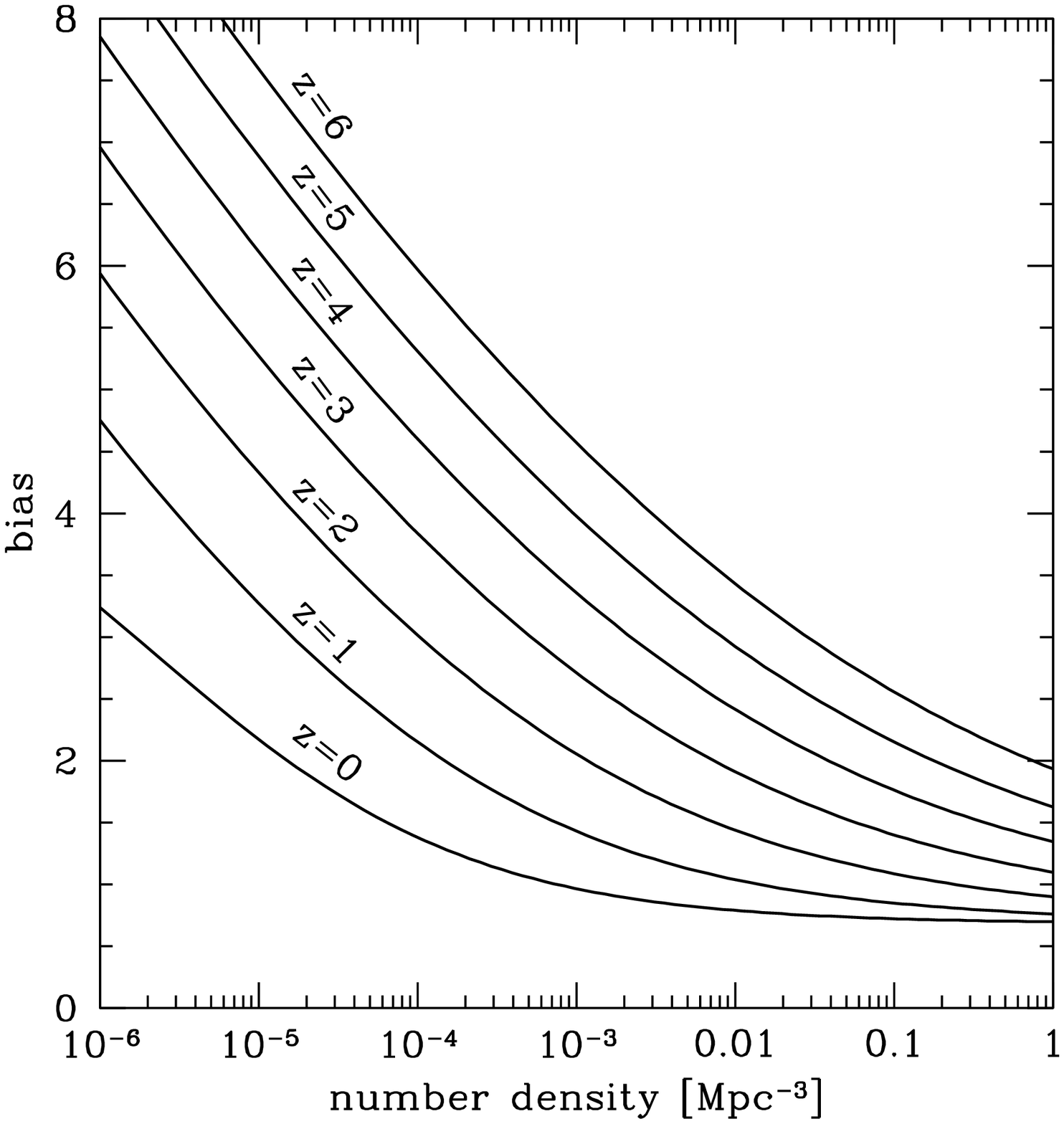}{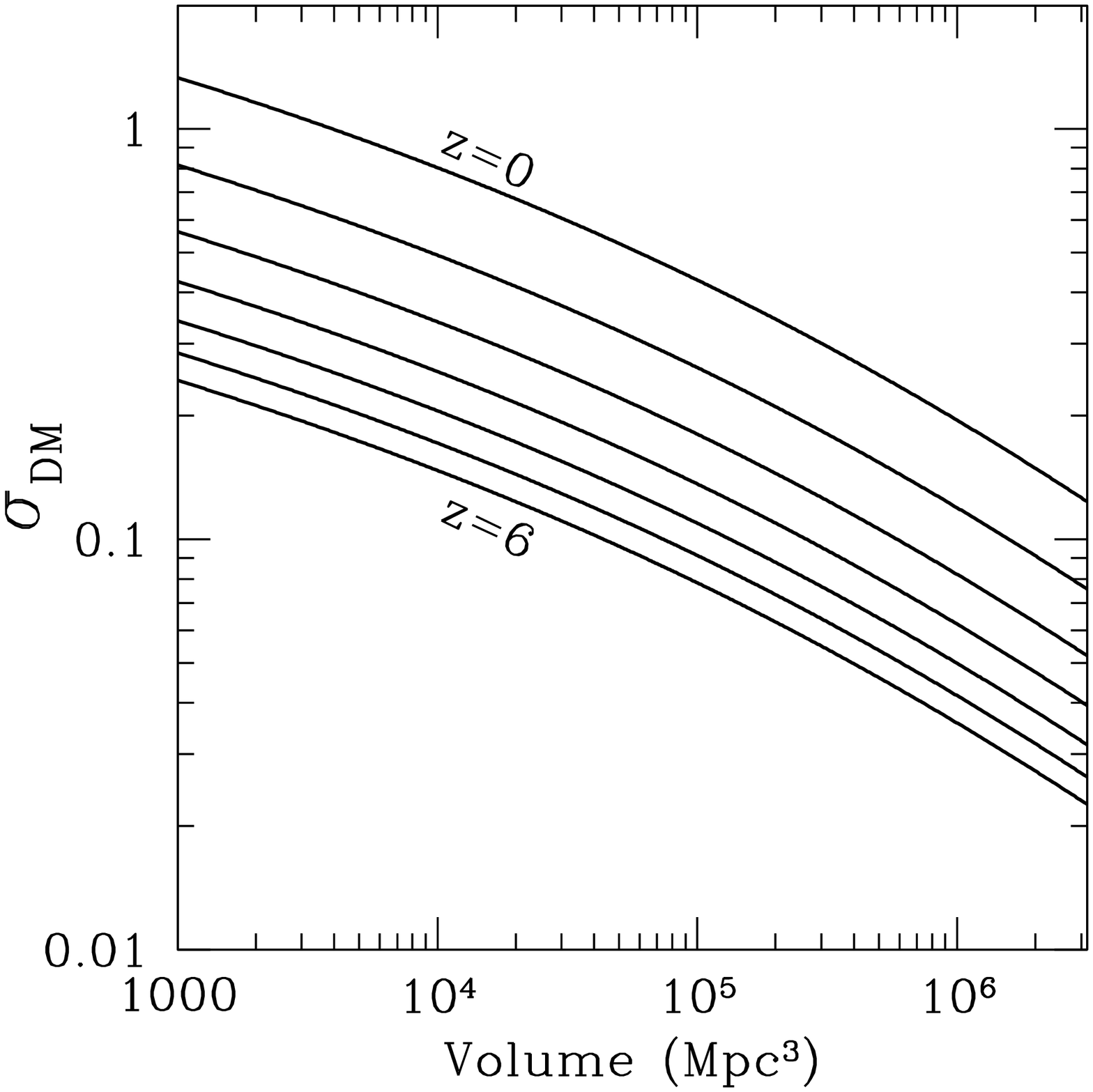}
\caption{\small [left] Bias as a function of comoving number density,
for dark matter halos at $z=6$, 5, 4, 3, 2, 1, and $z=0$ (from top to
bottom). [right] Variance of dark matter from linear theory, for the
same redshifts ($z=0$--6 from top to bottom).
\label{fig:bias}}
\end{figure*}
%\clearpage
We now consider the situation in which we know only the number density
but not the correlation function of a population. In this situation,
we can use predictions from CDM to estimate the clustering strength of
a population with a given number density at a known average
redshift. In CDM models, both the number density and clustering
strength of dark matter halos are a strong function of halo
mass. Fig.~\ref{fig:bias} shows the average bias as a function of
number density, for dark matter halos at various redshifts (as
described in the figure caption), computed using the analytic model of
\citet{st:99}. The bias is defined as the ratio of the root variance
of the halos and the dark matter, $b \equiv
\sigma_{h}/\sigma_{DM}$. It is likely that this relationship is more
complicated for galaxies, since there is probably not a one-to-one
correspondence between galaxies and dark matter halos. Similar
relations for more general `occupation functions' (i.e., allowing
varying numbers of galaxies per halo) are given in
e.g. \citet{ms:02}. Fig.~\ref{fig:bias} also shows the variance of
dark matter $\sigma_{DM}$ as a function of cell volume for the same
redshifts, as predicted by linear theory ($\sigma(R, z) = \sigma(R, 0)
\, D_{\rm lin}$, where $D_{\rm lin}$ is the linear growth
function). We now have all the ingredients necessary to obtain a rough
estimate of the cosmic variance for any population associated with
dark matter halos with a known number density and mean redshift:

\begin{enumerate}
\item Read off the average bias $b$ for objects of a given number
density and mean redshift from the left panel of Fig.~\ref{fig:bias}.
\item Obtain the value of $\sigma_{DM}$ at the relevant scale $V$ and
redshift from the right panel of Fig.~\ref{fig:bias}.
\item The cosmic variance for the population is then given by $\sigma_v =
b\,\sigma_{DM}$.
\end{enumerate}

As a consistency check, we can use the values given in
Table~\ref{tab:param} to estimate the cosmic variance for the same
populations discussed in the previous section. For EROs, using the
number density $n=1.0\times10^{-3} {\rm h^3} {\rm Mpc^{-3}}$, we would
estimate a bias of $b\sim1.8$, resulting in $\sigma_v \sim 0.7$ at
$z=1$, in reasonable agreement with our earlier estimate of $\sigma_v
\sim 0.6$ at $z=1.2$. Similarly, for LBGs at $z=3$, we find
$b\sim2.5$, resulting in $\sigma_v \sim 0.25$, again in agreement with
the earlier estimate of $\sigma_v \sim 0.2$. One reason that these
estimates are not in precise agreement with the values obtained from
the calculation based on the actual correlation length is due to the
unknown halo occupation distribution (i.e., the number of galaxies per
halo as a function of halo mass). It has been shown previously that
one cannot simultaneously exactly reproduce both the number density
and observed correlation length of either of these populations under
the simple assumption of one galaxy per halo \citep{ms:02}, adopted
here.

\section{Conclusions}

Cosmic variance can be a significant source of uncertainty in
estimates of the number density or related quantities in deep
surveys. We have given empirical estimates of the uncertainty due to
cosmic variance for several populations that have been identified in
the GOODS survey: EROs at $z\sim1$, U-dropouts at $z\sim3$ and
B-dropouts at $z\sim4$. These empirical estimates were based on
correlation function measurements from the literature for similarly
defined populations, and may be refined once correlation function
estimates have been obtained for the actual populations identified in
GOODS. From this calculation, we saw that the cosmic variance is much
reduced in GOODS compared with the original HDF (40--60\% rather than
a factor of 2 for for very strongly clustered populations such as
EROs, 15--20\% rather than 40\% for less clustered populations such as
LBGs). We have also presented predictions from the theory of
clustering and bias in a \LCDM\ Universe, which allow one to estimate
the cosmic variance for a population of a known average redshift and
number density but unknown clustering strength. We emphasize that this
approach is intended to give only a simple first order estimate of the
cosmic variance. More detailed estimates, tailored to individual
populations and including treatments of e.g. a generalized halo
occupation distribution formalism, geometric effects, the
observational selection function, and clustering evolution and the
change in absolute magnitude limit over the redshift range of the
sample, will be addressed in future works.

%===================================
\section*{Acknowledgments}
\begin{small}

We thank Emanuele Daddi and Mike Fall for useful discussions and
comments. Support for this work was provided by NASA through grant
GO09583.01-96A from the Space Telescope Science Institute, which is
operated by the Association of Universities for Research in Astronomy,
under NASA contract NAS5-26555.

\end{small}
%=====================================

\bibliographystyle{apj} 
\bibliography{apj-jour,cosvar}

%\clearpage

\end{document}